\documentclass[letter,twocolumn]{jpsj2} 
\title{Absence of Anomalous Tunneling of Bogoliubov Excitations for Arbitrary Potential Barrier under the Critical Condensate Current}

\author{Daisuke Takahashi\thanks{E-mail: takahashi@vortex.c.u-tokyo.ac.jp} and Yusuke Kato}

\inst{Department of Basic Science, University of Tokyo, Tokyo 153-8902}

\abst{We derive the exact solution of low energy limit of Bogoliubov equations for excitations of Bose-Einstein condensate in the presence of arbitrary potential barrier and maximum current of condensate. Using this solution, we give the explicit expression for the transmission coefficient against the potential barrier, which shows partial transmission in the low energy limit. The wavefunctions of excitations in the low energy limit do not coincide with that of the condensate. The absence of the perfect transmission in the critical current state originates from local enhancement of density fluctuations around the potential barrier.}

\kword{Bose-Einstein condensate, Gross-Pitaevskii equation, Bogoliubov equations, anomalous tunneling, critical current}
\begin{document}
\maketitle
\begin{figure}[tb]
		\begin{center}
		\begin{tabular}[c]{cc}
		\includegraphics[scale=0.8]{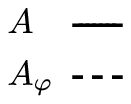}
		&
		\includegraphics[scale=0.8]{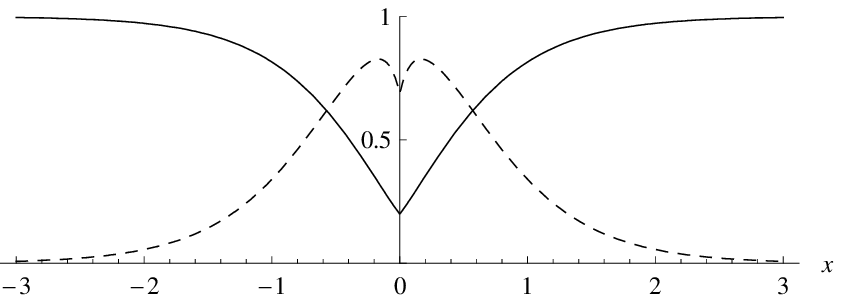}
		\end{tabular}
		\caption{\label{fig1} Localized density fluctuation solution $ A_\varphi=\partial A/\partial\varphi $ for a delta-functional barrier $ U(x)=V_0\delta(x) $ with $ V_0=3.32 $. (We multiply $ A_\varphi $ by a constant in the figure.)}
		\end{center}
	\end{figure}

	Anomalous tunneling, which was first discovered by Kovrizhin and his collaborators,\cite{kov} is an interesting theoretical prediction concerning the tunneling properties of Bogoliubov excitations of weakly-interacting Bose gas in superfluid phase; Bogoliubov excitations tunnel across a potential barrier without reflection in low energy limit. This prediction has a relevance to experiments in magnetically trapped dilute Bose-Einstein condensates, where the propagation of collective excitations\cite{Andrews} and the structure factor of density fluctuation\cite{Stamper,Steinhauer} have been accounted for by Bogoliubov theory\cite{Bogoliubov}.\\
	\indent Generalized or related issues of the anomalous tunneling have been considered by several authors\cite{danshita2,Bilas,danshita,danshita3,watabe}. Particularly, Danshita \textit{et~al.}\cite{danshita2} have considered the tunneling properties of excitations of Bose system in the presence of supercurrent of condensate, and have found that (a) the perfect transmission occurs even when the condensate current exists, except for the critical current state; (b)under the critical current, the perfect transmission disappears, and only a partial transmission occurs.\\
	 \indent Recently, the anomalous tunneling has been proved for an arbitrary barrier in the absence of condensate current \cite{kato}, using the fact that wavefunctions of Bogoliubov excitations in the low energy limit coincide with that of a condensate. Subsequently, Ohashi-Tsuchiya discussed the origin of anomalous tunneling considering the condensate current \cite{ohashi} and attributed (a) to the similarity between a low-energy Bogoliubov excitation and a condensate wavefunction.\\
	\indent From the scenario\cite{kato,ohashi}, the anomalous tunneling is expected to be a universal phenomenon related to excitations of superfluid condensate. Therefore the anomalous tunneling should be understood within the framework applicable to arbitrary shape of potential barrier. However, earlier works \cite{kov,danshita2,ohashi} but refs.~\citen{watabe,kato} have been based on the exact solutions for a particular shape of potential barrier and there is no discussion on (b) for general potential barrier so far. It is thus important to clarify whether the scenario \cite{kato,ohashi} of the origin of the anomalous tunneling explains consistently both the perfect transmission in the non-critical states and the partial transmission in the presence of critical condensate current for general potential barrier. \\
	\indent In this Letter, we derive the exact expression for wavefunctions of Bogoliubov excitations of Bose condensate under the critical supercurrent in the presence of potential barrier with an arbitrary shape. With use of this result, we prove the partial transmission (i.e., the absence of perfect transmission ) of excitations in the low energy limit under the critical current. The low energy limit of wavefunctions of excitations does not coincide with that of the condensate wavefunction, as a result of the emergence of low energy density fluctuations localized around the potential barrier.\\ 
	\indent Time-dependent Gross-Pitaevskii (GP) equation in dimensionless form is given by
	\begin{equation}
	\mathrm{i}\frac{\partial}{\partial t}\psi(x,t)=\left(\!-\frac{1}{2}\frac{\partial^2}{\partial x^2}\!+\!U(x)\!\right)\!\psi(x,t)+|\psi(x,t)|^2\psi(x,t).\nonumber
	\end{equation}
	One can always recover the dimensional form by defining $ x_{\text{dim}}=\xi x,\,t_{\text{dim}}=\hbar t/(gn_0),\,U_{\text{dim}}(x_{\text{dim}})=gn_0 U(x)$, and $ \psi_{\text{dim}}(x_{\text{dim}},t_{\text{dim}})=\!\sqrt{n_0}\psi(x,t) $, where $ g $ is the strength of repulsive interaction, $ n_0 $ is the condensate density far from the barrier, and  $ \xi = \hbar/\!\sqrt{mgn_0} $ is the healing length. Setting the condensate wavefunction in the form of
	\begin{equation}
	\psi(x,t)={\rm e}^{-\mathrm{i} \mu t}\!\left\{\Psi(x)+\left[u(x)\mathrm{e}^{-\mathrm{i} \epsilon t}-v^*(x)\,\mathrm{e}^{\mathrm{i} \epsilon t}\right]\right\}\label{eq: smallpsi}
	\end{equation}
	and taking the terms up to first order with respect to $u(x), v(x)$, we obtain the stationary GP equation
	\begin{gather}
		\hat{L}\Psi(x)=0,\ \hat{L} = -\frac{1}{2}\frac{\mathrm{d}^2 }{\mathrm{d} x^2} +U(x)-\mu+ |\Psi(x)|^2 \label{eqGP}
	\end{gather}
	for condensate wavefunction and Bogoliubov equations 
	\begin{gather}
		\begin{pmatrix}\! \hat{L}+|\Psi(x)|^2 & -(\Psi(x))^2 \! \\ \! -(\Psi(x)^*)^2 & \hat{L}+|\Psi(x)|^2 \! \end{pmatrix}\!\! \begin{pmatrix}\! u(x)\! \\ \!v(x)\! \end{pmatrix} = \epsilon \!\begin{pmatrix}\! u(x)\! \\ \!-v(x) \! \end{pmatrix}\! \label{eqBogo}
	\end{gather}
	for wavefunctions of excitations.\\
	\indent We assume that $ U(x) $ is short-ranged, and consider the solution which has the following asymptotic form: 
	\begin{align}
		\Psi(x\rightarrow\pm\infty) = \exp\Big[\mathrm{i}\Big(qx\pm\frac{\varphi}{2}+\text{const.}\Big)\Big]. \label{eqGPasymp}
	\end{align}
	The supercurrent $q$ depends on the phase difference $\varphi$ between two condensates separated by the potential barrier. The symbol ``const.'' represents a trivial non-uniqueness of the phase factor.\\
	\indent Setting $ \Psi(x) = A(x) \mathrm{e}^{\mathrm{i}\Theta(x)} $ in eq. (\ref{eqGP}), one obtains
	\begin{gather}
		-\frac{1}{2}\frac{\mathrm{d}^2 A}{\mathrm{d} x^2} + \frac{1}{2}\!\left( \frac{\mathrm{d} \Theta}{\mathrm{d} x} \right)^2\!A + \left( U-\mu \right)A+A^3=0, \label{eq: A-Theta}\\
		\frac{\mathrm{d} }{\mathrm{d} x} \!\left( A^2\frac{\mathrm{d} \Theta}{\mathrm{d} x} \right)\! = 0.\label{eq: Theta}
	\end{gather}
	From (\ref{eqGPasymp}), the boundary condition $ A(\pm\infty)=1 $ and the chemical potential $ \mu=1+q^2/2 $ follow. From (\ref{eqGPasymp}) and (\ref{eq: Theta}), we can obtain the expression for the phase of condensate
	\begin{gather}
		\!\Theta(x) = q\!\int_0^x\!\!\frac{\mathrm{d}x'}{A(x')^2} = qx + q\!\int_0^x\!\!\mathrm{d}x'\!\left( \frac{1}{A(x')^2}-1 \right)\!, \label{eqphase0}
	\end{gather}
	and the expression for the phase difference
	\begin{gather}
		\varphi = q\!\int_{-\infty}^{+\infty}\!\!\mathrm{d}x\!\left( \frac{1}{A(x)^2}-1 \right). \label{eqphase}
	\end{gather}
	Here we fix $\Theta(x\!=\!0)$ to be zero. By eliminating ${\rm d}\Theta/{\rm d}x$ in (\ref{eq: A-Theta}), we obtain
	\begin{gather}
		\hat{H}A=0,\ \hat{H} = -\frac{1}{2}\frac{\mathrm{d}^2 }{\mathrm{d}x^2}+U+\frac{q^2}{2\,}\Bigl(\frac{1}{A^4}\!-\!1 \Bigr)\!-1\!+\!A^2. \label{eqGPamp}
			\end{gather}
	Generally, eq. (\ref{eqGPamp}) has one or more solutions for given $ q $, so $ \varphi $ is more suitable than $ q $ as a parameter which specifies a state of the system uniquely. Therefore, we consider $ q $ and $ A $ as functions of $ \varphi $, i.e., $ q(\varphi) $ and $ A(x,\varphi) $. $q\,\text{-}\,\varphi$ curve corresponds to Josephson relation \cite{baratoff,danshita2}. An example is available in Fig.~2 of ref.~\citen{baratoff}.  The state where $ q $ takes an extremal value is the critical current state.\\
	\indent Introducing 
	\begin{align}
	S=u\mathrm{e}^{-\mathrm{i}\Theta}\!+v\mathrm{e}^{\mathrm{i}\Theta}, \quad G=u\mathrm{e}^{-\mathrm{i}\Theta}\!-v\mathrm{e}^{\mathrm{i}\Theta}, \label{eq: uv-SG}
	\end{align}
 eq. (\ref{eqBogo}) can be rewritten as 
	\begin{align}
		\hat{H}S-\frac{\mathrm{i}q}{A}\frac{\mathrm{d}}{\mathrm{d}x}\!\left( \frac{G}{A} \right)\! = \epsilon G, \\
		(\hat{H}+2A^2)G-\frac{\mathrm{i}q}{A}\frac{\mathrm{d}}{\mathrm{d}x}\!\left( \frac{S}{A} \right)\! = \epsilon S.
	\end{align}
	$ S $ and $ G $ have the simple physical meanings; $|\psi|^2$ and $\psi/|\psi|$ of (\ref{eq: smallpsi}) can be written, respectively, as
	\begin{gather}
		|\psi|^2 = A^2 \Bigl[1+\frac2A \operatorname{Re}(G{\rm e}^{-\mathrm{i}\epsilon t})\Bigr], \\
		\begin{split}
		\frac{\psi}{|\psi|}&={\rm e}^{-\mathrm{i}\mu t+\mathrm{i}\Theta}\Bigl[1+\frac{\mathrm{i}}{A}\operatorname{Im}(S{\rm e}^{-\mathrm{i}\epsilon t})\Bigr], 
		\end{split}
	\end{gather}
	where higher-order terms of $ S $ and $ G $ are ignored. From these expressions, we can regard $S/A$ and $G/A$ as the phase and density fluctuations, respectively. It has been also pointed out by Fetter and Rokhsar\cite{FetterRoksar}.\\
	\indent In the following, we consider the tunneling problem of excitations and hence we seek for the solution having the asymptotic form:
	\begin{align}
		S(x) \rightarrow \begin{cases} \mathrm{e}^{\mathrm{i}k_1x}+\tilde{r}\,\mathrm{e}^{\mathrm{i}k_2x} &(x\rightarrow-\infty) \\ t\,\mathrm{e}^{\mathrm{i}k_1x} &(x\rightarrow+\infty) 
		\label{eq: S(x) asym}
		\end{cases}.
	\end{align}
	$ k_1 $ and $ k_2 $ are real positive and negative roots of the dispersion relation $ \epsilon=qk+\frac{1}{2}\sqrt{k^2(k^2+4)} $, and $ k_1\simeq \epsilon/(1+q) $ and $ k_2\simeq \epsilon/(-1+q) $, respectively. It should be noted that  $ |t|^2 $ represents a transmission coefficient, while $ |\tilde{r}|^2 $ is not a reflection coefficient.\cite{danshita2}\\
	\indent We construct the solution of the Bogoliubov equations in the form of the power series in $ \epsilon $ :
	\begin{align}
		\begin{pmatrix}\! S(x) \! \\ \! G(x) \!\end{pmatrix} = \sum_{n=0}^\infty \epsilon^n \begin{pmatrix}\! S^{(n)}\!(x) \! \\ \! G^{(n)}\!(x) \!\end{pmatrix}.
	\end{align}
	Substituting it to the Bogoliubov equations, one obtains the recurrence relations:
	\begin{align}
		\hat{H}S^{(n)}-\frac{\mathrm{i}q}{A}\frac{\mathrm{d}}{\mathrm{d}x}\!\left( \frac{G^{(n)}}{A} \right)\! = G^{(n-1)}, \label{eqbogoS17}\\
		(\hat{H}+2A^2)G^{(n)}-\frac{\mathrm{i}q}{A}\frac{\mathrm{d}}{\mathrm{d}x}\!\left( \frac{S^{(n)}}{A} \right)\! = S^{(n-1)}. \label{eqbogoG17}
	\end{align}
	From these equations, we see that $S^{(n)}\!,G^{(n)}$ can be determined from $S^{(0)}\!,G^{(0)}$, recursively. The general solution of eqs. (\ref{eqbogoS17}) and (\ref{eqbogoG17}) should be expressed in terms of four homogeneous and one particular solution(s):
	\begin{align}
		\begin{pmatrix}\! S^{(n)}\!(x) \! \\ \! G^{(n)}\!(x) \!\end{pmatrix} &= \!\!\!\sum_{j=\text{I,II,III,IV}}\!\!\!\! C_j \begin{pmatrix}\! S_j(x) \! \\ \! G_j(x) \!\end{pmatrix}+\biggl(\begin{matrix} S^{(n)}_{\text{p}}\!(x) \\ G^{(n)}_{\text{p}}\!(x) \end{matrix}\biggr). \label{eqinhm}
	\end{align}
	Here the last term in the right hand side does not exist if $ n\!=\!0 $.\\
	\indent First, we look for the general solution for $ n\!=\!0 $, or equivalently, the homogeneous solutions $ (S_j,G_j) $. Since Bogoliubov equations always have the solution $ \epsilon=0 $ and $ (u,v)=(\Psi,\Psi^*) $,\cite{fetter} one can take  $ (S_{\text{I}},G_{\text{I}})=(A,0) $. So the solution of eq. (\ref{eqbogoS17}) for $ n\!=\!0 $ becomes
	\begin{align}
		S^{(0)} =&\  C_{\text{I}}A + C_{\text{II}}A\!\int_0^x\!\frac{\mathrm{d}x}{A^2}-2\mathrm{i}qA\!\int_0^x\!\frac{G^{(0)}\mathrm{d}x}{A^3}
		 \label{solbogoS}
	\end{align}
	by regarding $ G^{(0)} $ as an inhomogeneous term. Substituting (\ref{solbogoS}) to (\ref{eqbogoG17}) for $ n\!=\!0 $, we obtain the following equation for $ G^{(0)} $:
	\begin{align}
		\left(\!\hat{H}+2A^2-\frac{2q^2}{A^4} \right)\!G^{(0)}=\mathrm{i}qC_{\text{II}}A^{-3}. \label{eqbogoG2}
	\end{align}
	On the other hand, differentiating eq. (\ref{eqGPamp}) with respect to $ \varphi $, one obtains 
	\begin{align}
		\!\!\!\!\left(\!\hat{H}+2A^2-\frac{2q^2}{A^4}\!\right)\!A_\varphi \!= q\frac{\mathrm{d}q}{\mathrm{d}\varphi}(A\!-\!A^{-3}), \ A_\varphi:=\frac{\partial A}{\partial\varphi}. \label{eqGPphid2}
	\end{align}
	Comparing (\ref{eqbogoG2}) with (\ref{eqGPphid2}), we find that $ A_{\varphi} $ becomes a homogeneous solution of (\ref{eqbogoG2}), \textit{only when the condition $ \mathrm{d}q/\mathrm{d}\varphi=0 $ of critical current holds}. Since $ A(x\!\rightarrow\!\pm\infty)=1 $, $ A_\varphi(x\!\rightarrow\!\pm\infty)=0$, i.e., $A_\varphi(x)$ is \textit{localized} near the potential barrier. As an example, see Fig.~\ref{fig1}.
	Once we know this solution, other solutions can be obtained by methods of reduction of order and variation of parameters. We thus obtain all homogeneous solutions as follows:\vspace{-0.5ex}
	\begin{gather*}
		\begin{pmatrix}\! S_{\text{I}} \! \\ \! G_{\text{I}} \!\end{pmatrix} \!=\! \begin{pmatrix} A \\ 0 \end{pmatrix}\!, \ \ \begin{pmatrix}\! S_{\text{II}} \!\\ \!G_{\text{II}} \!\end{pmatrix} \!=\! \begin{pmatrix}\displaystyle \! A\!\int^x_0\!\frac{\mathrm{d}x}{A^2}-2\mathrm{i}qA\!\int^x_0\!\frac{G_{\text{II}}\mathrm{d}x}{A^3} \\[2ex] \displaystyle -2\mathrm{i}qA_\varphi\!\int^x_0\!\frac{A_3\mathrm{d}x}{A_\varphi^2} \end{pmatrix}\!, \\
		\begin{pmatrix}\! S_{\text{III}} \! \\ \! G_{\text{III}} \!\end{pmatrix} \!=\! \begin{pmatrix}\displaystyle -2\mathrm{i}qAA_3 \\ \displaystyle A_{\varphi} \end{pmatrix}\!, \ \ \begin{pmatrix}\! S_{\text{IV}} \! \\ \! G_{\text{IV}} \!\end{pmatrix} \!=\! \begin{pmatrix}\displaystyle \!-2\mathrm{i}qA\!\int^x_0\!\frac{G_{\text{IV}}\mathrm{d}x}{A^3} \\[2ex] \displaystyle A_\varphi\!\int^x_0\!\frac{\mathrm{d}x}{A_\varphi^2} \end{pmatrix}\!.
	\end{gather*}
	Here we have introduced the following function:\vspace{-0.5ex}
	\begin{align}
		A_3(x) := \int_0^x\!\frac{A_{\varphi}(x')\mathrm{d}x'}{A(x')^3}. \label{eqa3}
	\end{align}
	We note that the third solution $ (S_{\text{III}},G_{\text{III}}) $ can be expressed as $ (u,v) = (\partial\Psi/\partial\varphi,-\partial\Psi^*/\partial\varphi)$ by means of (\ref{eqphase0}) and (\ref{eq: uv-SG}). We also note that $ (S_{\text{II}},G_{\text{II}}) $ and $ (S_{\text{IV}},G_{\text{IV}}) $ are exponentially divergent solutions.\\
	\indent Since we know all homogeneous solutions, the particular solution $ (S^{(n)}_{\text{p}}\!,G^{(n)}_{\text{p}}) $ in (\ref{eqinhm}) can be obtained by a method of variation of parameters. It is
	\begin{gather}
	\begin{align}
		\!\!\! G^{(n)}_{\text{p}}\!(x)=& -\!2A_\varphi(x)\!\!\int_0^x\!\!\!\frac{\mathrm{d}x'}{A_\varphi(x')^2}\!\int_0^{x'}\!\!\!\!\mathrm{d}x''A_\varphi(x'')\biggl[S^{(n-1)}\!(x'')\nonumber \\[-1ex]
		&\ \  -\frac{2\mathrm{i}q}{A(x'')^3}\!\!\int_0^{x''}\!\!\!\mathrm{d}x'''A(x''')G^{(n-1)}\!(x''')\biggr],
	\end{align}\\[0ex]
	\begin{align}
		\!\!\! S^{(n)}_{\text{p}}\!(x)=& -\!2A(x)\!\int_0^x\!\frac{\mathrm{d}x'}{A(x')^2}\!\int_0^{x'}\!\!\mathrm{d}x''A(x'')G^{(n-1)}\!(x'') \nonumber \\[-1ex]
		& \ \ -2\mathrm{i}qA(x)\!\int_0^x\!\frac{G^{(n)}_{\text{p}}\!(x')\mathrm{d}x'}{A(x')^3}.
	\end{align}
	\end{gather}
	\indent From now on, we assume that the barrier $ U(x) $ is even for simplicity. Accordingly, $ A(x) $ and $ A_\varphi(x) $ are also even. The asymptotic behavior of $ A_\varphi(x) $ is 
	\begin{align}
		A_\varphi(x) \rightarrow \beta\,\mathrm{e}^{-2\sqrt{1-q^2}|x|} \quad (x\rightarrow\pm\infty), \label{Basym}
	\end{align}
	where $ \beta $ is a certain constant. This asymptotics comes from the solution of (\ref{eqbogoG2}) far from the barrier ($ U(x)\simeq 0, A(x)\simeq 1 $).\\
	\indent In the tunneling problem, we are interested in solutions free from exponential divergence. Accordingly, $ (S^{(n)}\!,G^{(n)}) $ must be chosen to behave asymptotically as a polynomial of finite order in $ x $. In general,  $ (S^{(n)}_{\text{p}}\!,G^{(n)}_{\text{p}}) $ itself diverges exponentially. Nevertheless we can construct a particular solution that does not diverge exponentially by making a linear combination with $ (S_{\text{II}},G_{\text{II}}) $ or $ (S_{\text{IV}},G_{\text{IV}}) $.\\
	\indent Now, we find $ (S^{(1)}\!,G^{(1)}) $ when a zero-energy solution free from divergence is chosen to be $ (S^{(0)}\!,G^{(0)}) = (S_{\text{I}},G_{\text{I}}) $ or $ (S_{\text{III}},G_{\text{III}}) $. For this purpose, we introduce the notations $ A_1(x) := \int_0^x\!\!\mathrm{d}x'A(x')A_\varphi(x'),\,\alpha_1 := A_1(+\infty),\,\alpha_3 := A_3(+\infty) $, and $ \eta := \alpha_1/\alpha_3 $. Both $ A_1(x) $ and $ A_3(x) $ are odd functions, and from eq. (\ref{Basym}) they behave as 
	\begin{align}
		A_i(x) \rightarrow \operatorname{sgn}x\biggl(\!\alpha_i-\frac{\beta}{2\sqrt{1\!-\!q^2}}\mathrm{e}^{-2\sqrt{1-q^2}|x|} \biggr)
	\end{align}
	with $i=1,3$. We note that $ \alpha_3 $ can be simplified by means of eq. (\ref{eqphase}): $ \alpha_3\!=\!-\frac{1}{2}\frac{\partial}{\partial \varphi}\int_0^\infty\mathrm{d}x\,(A^{-2}\!-\!1)\!=\!-\frac{1}{4q} $. However, we hold this notation for a moment.\\
	\indent When $ (S^{(0)},G^{(0)}) =(S_{\text{I}},G_{\text{I}})=(A,0) $, the solution that does not diverge exponentially is given by
	\begin{align}
	\begin{split}
		&\biggl(\begin{matrix} S^{(1)}_{\text{I}}\!(x) \\ G^{(1)}_{\text{I}}\!(x) \end{matrix}\biggr) := \biggl(\begin{matrix} S^{(1)}_{\text{p}}\!(x) \\ G^{(1)}_{\text{p}}\!(x) \end{matrix}\biggr)-\frac{\eta}{\mathrm{i}q}\begin{pmatrix}\! S_{\text{II}}(x) \!\\ \!G_{\text{II}}(x) \!\end{pmatrix} \\[-0.5ex]
		&= \begin{pmatrix} \displaystyle -\frac{\eta}{\mathrm{i}q}A\!\int_0^x\!\frac{\mathrm{d}x}{A^2}-2\mathrm{i}qA\!\int_0^x\!\frac{G^{(1)}_{\text{I}}\mathrm{d}x}{A^3} \\[2ex] \displaystyle -2A_\varphi\!\int_0^x\!\!\mathrm{d}x\frac{A_1-\eta A_3}{A_\varphi^2} \end{pmatrix} \\[-0.5ex]
		&\!\overset{x\rightarrow\pm\infty}{\longrightarrow} \!\!\!\begin{pmatrix}\displaystyle \!-\frac{\eta}{\mathrm{i}q}(x\!+\!\gamma\operatorname{sgn}x)-\frac{\mathrm{i}q(1-\eta)}{1-q^2}\left(x\!+\!\nu\operatorname{sgn}x\right) \! \\[1.5ex] (1-\eta)/\bigl(2(1-q^2)\bigr) \end{pmatrix}\!\!,
	\end{split}
	\end{align}
	where  $ \gamma $ and $ \nu $ are certain constants. When $ (S^{(0)},G^{(0)}) =(S_{\text{III}},G_{\text{III}})=(-2\mathrm{i}qAA_3,A_\varphi) $, on the other hand, we obtain 
	\begin{align}
	\begin{split}
		&\biggl(\begin{matrix} S^{(1)}_{\text{III}}\!(x) \\ G^{(1)}_{\text{III}}\!(x) \end{matrix}\biggr) := \biggl(\begin{matrix} S^{(1)}_{\text{p}}\!(x) \\ G^{(1)}_{\text{p}}\!(x) \end{matrix}\biggr)-4\mathrm{i}q\alpha_1\alpha_3\begin{pmatrix}\! S_{\text{IV}}(x) \!\\ \!G_{\text{IV}}(x) \!\end{pmatrix} \\[-0.5ex]
		&= \begin{pmatrix} \displaystyle -2A\!\int_0^x\!\frac{A_1\mathrm{d}x}{A^2}-2\mathrm{i}qA\!\int_0^x\!\frac{G^{(1)}_{\text{III}}\mathrm{d}x}{A^3} \\[2ex] \displaystyle  4\mathrm{i}qA_\varphi\!\int_0^x\!\!\mathrm{d}x\frac{A_1A_3-\alpha_1\alpha_3}{A_\varphi^2} \end{pmatrix} \\[-0.5ex]
		&\!\overset{x\rightarrow\pm\infty}{\longrightarrow} \!\! \alpha_3\!\begin{pmatrix}\displaystyle -2\eta(|x|\!+\!\lambda)-\frac{2q^2(1+\eta)}{1-q^2}\left(|x|\!+\!\kappa\right)  \\[1.5ex] \displaystyle -\bigl(\mathrm{i}q(1+\eta)\bigr)/(1-q^2)\operatorname{sgn}x \end{pmatrix}\!\!,
	\end{split}
	\end{align}
	where  $ \lambda $ and $ \kappa $ are certain constants. Thus, non-divergent solutions up to first order in $ \epsilon $ are given by
	\begin{gather}
		\biggl(\begin{matrix} S^{\text{total}}_{i}(x) \\ G^{\text{total}}_{i}(x) \end{matrix}\biggr) := \begin{pmatrix}\! S_{i}(x) \! \\ \! G_{i}(x) \!\end{pmatrix} +\epsilon \biggl(\begin{matrix} S^{(1)}_{i}\!(x) \\ G^{(1)}_{i}\!(x) \end{matrix}\biggr) +O(\epsilon^2), 
	\end{gather}
	with $i={\rm I},{\rm III}$, and their asymptotic forms at $ x\rightarrow\pm\infty $ become \vspace{-0.75ex}
	\begin{gather}
		S^{\text{total}}_{\text{I}}(x) \rightarrow 1+\epsilon \left( \frac{q^2-\eta}{\mathrm{i}q(1-q^2)}x+\tilde{\gamma}\operatorname{sgn}x \right), \label{nondiv1}\\
		\frac{S^{\text{total}}_{\text{III}}(x)}{-2\mathrm{i}q\alpha_3} \rightarrow \operatorname{sgn}x +\epsilon\left( \frac{q^2+\eta}{\mathrm{i}q(1-q^2)}|x|+\tilde{\lambda} \right), \label{nondiv2}\\
		\tilde{\gamma}=-\frac{\eta}{\mathrm{i}q}\gamma-\frac{\mathrm{i}q(1-\eta)}{1-q^2}\nu,\,\ \tilde{\lambda}=\frac{\eta}{\mathrm{i}q}\lambda-\frac{\mathrm{i}q(1+\eta)}{1-q^2}\kappa.
	\end{gather}
	\indent With the above solutions, we derive the transmission coefficient $t$ in (\ref{eq: S(x) asym}). We expand (\ref{eq: S(x) asym}) in $ 1\ll |x|\ll 1/k_1 $, and expand coefficients by $ \epsilon $ such as $ t=t^{(0)}+\epsilon\,t^{(1)}+\dotsb,\,\tilde{r}=\tilde{r}^{(0)}+\epsilon\,\tilde{r}^{(1)}+\dotsb $, and we obtain
	\begin{align}
		\begin{cases} 1+\tilde{r}^{(0)}+\epsilon\left( \tilde{r}^{(1)}+\left( \frac{\mathrm{i}}{1+q}+\frac{\hphantom{{}^{(0)}}\mathrm{i}\tilde{r}^{(0)}}{-1+q} \right)x \right)+O(\epsilon^2) \\[2ex] t^{(0)}+\epsilon\left( t^{(1)}+\frac{\hphantom{{}^{(0)}}\mathrm{i}t^{(0)}}{1+q}x \right)+O(\epsilon^2) \end{cases}\!\!\!\!\!\!\!\!.
	\end{align}
	We then equate this expression with the asymptotic form of the general linear combination of solutions (\ref{nondiv1}) and (\ref{nondiv2}), that is,\vspace{-0.75ex}
	\begin{align}
		S(x) = C^{}_{\text{I}}S^{\text{total}}_{\text{I}}(x)+C^{}_{\text{III}}\frac{S^{\text{total}}_{\text{III}}(x)}{-2\mathrm{i}q\alpha_3}.
		\label{eq: S-of-x-final}
	\end{align}
	Expanding  $ C^{\mathstrut}_{\text{I}} $ and $ C^{\mathstrut}_{\text{III}} $ as $ C^{\mathstrut}_{\text{I}}=C_{\text{I}}^{(0)}+\epsilon\,C_{\text{I}}^{(1)}+\dotsb,\,C^{\mathstrut}_{\text{III}}=C_{\text{III}}^{(0)}+\epsilon\,C_{\text{III}}^{(1)}+\dotsb $, we can obtain simultaneous equations, and the solutions can be given by \vspace{-0.5ex}
	\begin{gather}
		t^{(0)} = \frac{2q\eta}{q^2+\eta^2},\ \tilde{r}^{(0)} = \frac{q^2-\eta^2}{q^2+\eta^2}, \\[-0.5ex]
		C_{\text{I}}^{(0)} = \frac{q(q+\eta)}{q^2+\eta^2},\ C_{\text{III}}^{(0)} = -\frac{q(q-\eta)}{q^2+\eta^2}. \label{eq: S-of-x-final2}
	\end{gather}
	It is obvious that $ 0<|t^{(0)}|^2<1 $ holds unless $ \eta=\pm q $, thus the partial transmission in the low energy limit follows. We note that $ S(x) $ must be expanded up to second order to obtain $ t^{(1)} $ and $ \tilde{r}^{(1)} $.\\
	\indent Let us make sure that our expression reproduces the result in ref. \citen{danshita2} for the delta-functional barrier  $ U(x)\!=\!V_0\delta(x) $. The condensate wavefunction is given by\cite{danshita2} $A(x)^2=\gamma(x)^2+q^2,\,\gamma(x)=\sqrt{1\!-\!q^2}\tanh\bigl(\sqrt{1\!-\!q^2}(|x|\!+\!x_0)\bigr)$. 
	Since $ \mathrm{d}q/\mathrm{d}\varphi=0 $ at the critical point, $ q $ can be regarded as a constant for differentiation with respect to $ \varphi $. Accordingly, $ A\,\partial A/\partial\varphi=\gamma\,\partial \gamma/\partial \varphi=\gamma\,(1\!-\!q^2)\partial x_0/\partial \varphi\,\cosh^{\!-2}\!\bigl(\!\sqrt{1\!-\!q^2}(|x|\!+\!x_0)\bigr)$. Using this expression, we can obtain $ \eta=q^2+\gamma(0)^2=A(0)^2 $. From ref.~\citen{danshita2}, when $ V_0\!\gg\!1 $, $ q_{\text{c}}\!\simeq\!\gamma(0)\!\simeq\!\frac{1}{2V_0} $, so $ \eta\!\simeq\!\frac{1}{2V_0{}^2} $. Therefore $ |t^{(0)}|^2\!\simeq\!\frac{4}{V_0{}^2} $, which is consistent with ref. \citen{danshita2}. Though ref. \citen{danshita2} gives the explicit expression only for the high barrier case, we have confirmed that our expression is exact irrespective of the height of barrier. We have also confirmed that $ \eta\!=\!q $ occurs only when  $ V_0\!=\!0 $, i.e., no barrier exists. Therefore, we expect that $ \eta\!=\!\pm q $ does not occur for a generic potential barrier, though we do not have the general proof for this criterion yet.\\
	\indent Finally, we discuss the physical origin of the disappearance of anomalous tunneling under the critical current. Recalling that $ (S_{\text{I}},G_{\text{I}}) $ and $ (S_{\text{III}},G_{\text{III}}) $ correspond to $ (u,v)=(\Psi,\Psi^*)\text{ and }(\partial\Psi/\partial\varphi,-\partial\Psi^*/\partial\varphi) $, respectively, and using (\ref{eq: S-of-x-final}) and (\ref{eq: S-of-x-final2}), we obtain
	\begin{align}
		\lim_{\epsilon\rightarrow 0}\begin{pmatrix} u \\ v \end{pmatrix} \propto \begin{pmatrix} \Psi^{\hphantom{*}}\! \\ \Psi^*\! \end{pmatrix} -2\mathrm{i}\,\frac{q\!-\!\eta}{q\!+\!\eta}\,\frac{\partial }{\partial\varphi}\!\!\begin{pmatrix}\Psi^{\hphantom{*}}\! \\ -\Psi^*\! \end{pmatrix},
	\end{align}
	i.e.,  $ \lim_{\epsilon\rightarrow 0} (u,v) \ne (\Psi,\Psi^*) $. This is in contrast to the case without supercurrent, in which the wavefunctions of excitations coincide with the condensate wavefunction.\cite{kato} 
	In the absence of supercurrent, the differential equation for $G^{(0)}$ has the unphysical exponentially divergent solutions only, and $G^{(0)}$ cannot contribute to the wavefunction of excitations in the low energy limit. 
	In the present case, on the other hand, the solution $G_{\rm III}\!=\!A_\varphi$, which is localized near the potential barrier, contributes to the low energy wavefunction of excitations. Thus the presence of local density fluctuation near the barrier in the low energy limit is the origin of the absence of anomalous tunneling in the critical current state. \\
	\indent We would like to thank I.~Danshita, S. Watabe and Y. Nagai for helpful discussions. This research was partially supported by the Ministry of Education, Science, Sports and Culture, Grant-in-Aid for Scientific Research on Priority Areas, 20029007.

\end{document}